\documentclass[twocolumn]{aastex63}
\usepackage{amsmath,amsthm,amssymb}

\shortauthors{Ghosh et al.}

\begin{document}

\title{Relativistic correction to the r-mode frequency in light of multi-messenger constraints}

\correspondingauthor{Suprovo Ghosh}
\email{suprovo@iucaa.in}

\author[0000-0002-1656-9870]{Suprovo Ghosh}
\affiliation{Inter-University Centre for Astronomy and Astrophysics, \\
Pune University Campus, Pune - 411007, India \\}

\author[0000-0001-8129-0473]{Dhruv Pathak}
\affiliation{Inter-University Centre for Astronomy and Astrophysics, \\
Pune University Campus, Pune - 411007, India \\}


\author[0000-0002-0995-2329]{Debarati Chatterjee}
\affiliation{Inter-University Centre for Astronomy and Astrophysics, \\
Pune University Campus, Pune - 411007, India \\}



\begin{abstract}

R-mode oscillations of rotating neutron stars are promising candidates for continuous gravitational wave (GW) observations. The r-mode frequencies for slowly rotating Newtonian stars are well-known and independent of the equation of state (EOS) but for neutron stars, several mechanisms can alter the r-mode frequency of which the relativistic correction is dominant and relevant for most of the neutron stars. The most sensitive searches for continuous GWs are those for known pulsars for which GW frequencies are in targeted narrow frequency bands of few Hz. In this study, we investigate the effect of several state-of-the-art multi-messenger constraints on the r-mode frequency for relativistic, slowly rotating, barotropic stars. Imposing these recent constraints on the EOS, we find that the r-mode frequency range is slightly higher from the previous study and the narrow band frequency range can increase upto 8-25\% for the most promising candidate PSR J0537-6910 depending on the range of compactness. We also derive universal relations between r-mode frequency and dimensionless tidal deformability which can be used to estimate the dynamical tide of the r-mode resonant excitation during the inspiral signal. These results can be used to construct the parameter space for r-mode searches in gravitational wave data and also constrain the nuclear equation of state following a successful r-mode detection.

\end{abstract}



\section{Introduction} \label{sec:Intro}
Gravitational waves (GW) can drive  various modes of oscillation in a rotating Neutron Star (NS) unstable ~\citep{CFS1,CFS2}. The r-mode is a toroidal mode of fluid oscillation for which
the restoring force is the Coriolis force. For any rotating star, r-modes can become unstable to gravitational wave emission via the Chandrasekhar-Friedman-Schutz (CFS)  mechanism~\citep{Andersson1998,Andersson2003}. This instability can explain the spin down of hot and young neutron stars~\citep{Andersson1999,Lindblom1998,Alford2014}
as well as old, accreting neutron stars in low mass x-ray
binaries (LMXBs)~\citep{Wynn2011}, and provides a plausible explanation
for the absence of very fast rotating neutron stars in nature. Although shear and bulk viscosity of the NS matter can damp these oscillation modes~\citep{Lindblomvis2002,Debi2006,Debi2007}, while spinning down, neutron stars can enter a region of temperature and rotational frequency in  which viscosity cannot damp the r-mode and
the mode grows to a large amplitude leading to GW emission~\citep{Lindblom1998}.\\

Because of its astrophysical significance~\citep{CWmili2022}, there have been recent searches for continuous GW emission specifically from r-modes using the LIGO-Virgo-Kagra (LVK) global network of gravitational wave detectors~\citep{AdvLIGO2015,Abbott2016,AdvVIRGO2014,Kagra}.  A search for GWs from r-modes in the Crab pulsar was carried out by ~\cite{Rajbhandari2021} and from PSR J0537 by ~\cite{Fesik2020a,Fesik2020b} and the LVK collaboration~\citep{LIGO_rmode}. No GWs were detected in these searches, but upper limits on GW amplitude were obtained.  \\

As the r-modes are rotationally restored stellar oscillations, their frequency is proportional to the rotation frequency~\citep{rmode1977} for slowly rotating stars. If the targeted pulsar rotational  frequency is known from electromagnetic data, the GW searches are done in a relatively narrow frequency band obtained from theoretical estimates~\citep{Caride2019}. The r-mode frequency for slowly and uniformly rotating Newtonian star in the perfect fluid approximation is found to be independent of the Equation of state (EOS)~\citep{Provost1981}
\begin{equation}\label{eq:freq}
    \kappa \equiv \frac{\sigma_R}{\Omega} = \frac{2m}{l(l+1)},
\end{equation}
where $\sigma_R$ is the r-mode frequency in the co-rotating frame, $\Omega$ is the rotational angular velocity of the star, $l$ and $m$ are the spherical harmonic indices. In the inertial frame, the frequency is given by $f = |(\kappa - m)\Omega|$. For neutron stars, dominant factors influencing the r-mode frequency are relativistic effects~\citep{Lockitch2001,Lockitch2003,idrisy2015} and rapid rotation~\citep{Lindblom1999,yoshida2005}. There are several other factors like the presence of solid crust~\citep{Yuri2001}, stratification~\citep{Yoshida2000,Passamonti2009}, magnetic field~\citep{Ho2000,Morsink2002} or superfluidity in the core~\citep{Lindblom2000,Andersson2001} that might affect the r-mode frequency but their effect was found to be negligible for most stars~\citep{idrisy2015}. Rapid rotation can increase the value of $\kappa$ by $\sim 6\%$ for fastest rotating stars~\citep{idrisy2015}. But since this correction is of the order $(\Omega/\Omega_K)^2$ where $\Omega_K$ is the Keplerian frequency, the rotational corrections become negligible for slowly spinning stars.  ~\cite{Lockitch2001,Lockitch2003} first derived the perturbation equations to solve for r-mode frequency for relativistic barotropic stars and calculated them using only polytropic EOSs. Later ~\cite{idrisy2015} calculated the same for some tabulated EOSs. They sampled 14 EOSs that support a maximum mass  of at least 1.85$M_{\odot}$ and obtained the  universal relation between r-mode frequency and compactness (Fig 3 in ~\citep{idrisy2015}) as 
\begin{equation}\label{universal}
    \kappa = 0.627 + 0.079(M/R) - 2.25(M/R)^2~.
\end{equation} 
Since this work, there have been several recent multi-messenger observations of neutron stars that have put constraints on the Neutron star equation of state. The detection of gravitational waves from the binary neutron star merger GW170817~\citep{Abbott2017,Abbott2018,Abbott2019,Abbott_2020}, the NICER measurement of mass and radius of pulsars~\citep{NICER0030_Miller,NICER0030_Riley,NICER0740_Miller,NICER0740_Riley} has been extensively used to constrain the nuclear equation of state~\citep{Pang_2021,Traversi2020,Dietrich2020,Legred2021,Annala,Biswas2021,Ghosh2022}. The maximum observed mass of pulsar is also updated to around $2.08 M_{\odot}$~\citep{fonseca2021refined}. These multi-messenger constraints~\citep{Abbott_2020} have ruled out with good confidence several tabulated EOSs that were used in ~\cite{idrisy2015}. Also, in the universal relation Eqn.~\ref{universal}, they did not impose two important physical constraints on the r-mode frequency that(i) in the Newtonian limit when $M/R \longrightarrow 0$, then $\kappa = 2/3(0.667)$ for $l = m =2$ mode and (ii) the linear coefficient in the fit relation should be negative, otherwise it implies that the r-mode frequency increases with increase in compactness upto a certain value~\citep{Lockitch2001}. \\

In this study, we calculate the r-mode frequency for 15 tabulated equations of state sampled from ~\citep{idrisy2015,Abbott_2020} that are consistent with recent multi-messenger observations of neutron stars. Along with these tabulated EOSs, we also consider a posterior for the EOS from~\cite{Legred2021}, employing a nonparametric EOS model based on Gaussian processes and combining information from pulsar masses, NICER observation of mass radius and GW observations of  binary neutron star mergers. The advantages of this nonparametric EOS model is that this allows more model freedom in the EOS representation than any direct parametrization with a small number of parameters; can account for different degrees of freedom, including hyperonic, quark models, phase transition and is not subject to any systematic errors that arise with parametrized EOS families~\citep{Legred2021}. A couple of recent studies~\citep{pawan2022,Ma2021} have also looked into the possible resonant r-modes detection in the inspiral phase of the binary mergers using the third generation detector Einstein telescope~\citep{ET1,ET2}. The r-mode introduces an additional phase to the waveform model which is estimated from the r-mode frequency $\kappa$ as function of the dimensionless tidal deformability of the neutron star. So, for both of these sets of EOS models, we  calculate the r-mode frequency as a function of dimensionless tidal deformability.  We also give universal relations of the r-mode frequency with both the compactness and dimensionless tidal deformability which will be useful to constrain the parameter space for searches for r-modes from both isolated pulsars and excitation during inspiral phase in binary. In case, gravitational wave is detected from the r-mode oscillations, these universal relations can also be used to constrain the neutron star EOS. \\

The structure of the article is as follows: in Sec.~\ref{sec:formalism}, we describe the formalism of the structure of equilibrium configuration of slowly and uniformly rotating star, relevant perturbation and boundary equations. The details of the numerical scheme and their convergence are discussed in Sec.~\ref{sec:numerical}. We first test the scheme by comparing with the results of \cite{idrisy2015,Lockitch2003} in Sec~\ref{sec:results}. We then extend the analysis for our EOS models and  demonstrate the results of the investigation, including the universal relations in Sec.~\ref{sec:results}. Finally, in Sec.~\ref{sec:discussion} we discuss the main implications of this work.

\section{Formalism}
\label{sec:formalism}
\subsection{Equilibrium configuration of slowly and uniformly rotating star}
We consider a slowly and uniformly rotating perfect fluid star with angular velocity $\Omega$. The \textit{slow rotation} approximation requires that $\Omega \ll \Omega_K$ where $\Omega_K$ is the Keplerian frequency($\propto \sqrt{M/R^3}$). In this \textit{slow-rotation} approximation, the neutron star retains its spherical geometry as the centrifugal deformations are an order $\Omega^2$ effect~\citep{Hartle}. The equilibrium solution to the slowly and uniformly rotating star is obtained by solving the Einstein equations $G_{\alpha\beta} = 8\pi T_{\alpha\beta}$ using the line element given by~\citep{Lockitch2001}
\begin{equation}\label{eq:metric}
\begin{split}
    ds^2 & =-e^{2\nu(r)} dt^2 + e^{2\lambda(r)} dr^2 + r^2 d\theta^2  \\ & \quad + r^2 \sin^2(\theta)d\phi^2 - 2\omega (r)r^2 \sin^2(\theta) dt d\phi ,
\end{split}   
\end{equation}

where $\omega (r)$ is the rotational frame-dragging  inside the star.
The energy momentum tensor for the perfect fluid is given by
\begin{equation}\label{energy}
    T_{\alpha\beta} = (\varepsilon + p)u_{\alpha}u_{\beta} + pg_{\alpha\beta}.
\end{equation}
Here, $\varepsilon,p$ are  the total energy density and pressure of the fluid respectively measured by an observer co-moving with the 4-velocity
\begin{equation}
    u^{\alpha} = e^{-\nu}(t^{\alpha} + \Omega \phi^{\alpha}),
\end{equation}
where $t^{\alpha}$ and $\phi^{\alpha}$ are the timelike and rotational killing vectors of the spacetime.
 Solving the Einstein equations for the metric and fluid variables, they reduce to  Tolman-Oppenheimer-Volkov (TOV) equations given in Eqn.~\eqref{eq:tov}
\begin{eqnarray}
\frac{dm(r)}{dr} &=& 4 \pi \varepsilon(r) r^2 ~, \nonumber \\
\frac{dp(r)}{dr} &=& - \frac{[p(r) + \varepsilon(r)] [m(r)+4 \pi r^3 p(r)] }{r(r-2 m(r))}.~ \nonumber \\
\label{eq:tov}
\end{eqnarray}
We also obtain the equations for the metric functions $\nu(r)$ and $\lambda(r)$ as
\begin{eqnarray}
\frac{d\nu(r)}{dr} &=& -\frac{1}{\varepsilon + p } \frac{dp(r)}{dr}, \nonumber \\
e^{-2\lambda} &=& 1 - \frac{2m(r)}{r}~. \nonumber \\
\label{eq:metricf}
\end{eqnarray}
For a given EOS $p = p(\varepsilon)$, the TOV equations~\eqref{eq:tov} are integrated from the centre of the star to the surface using the boundary conditions of vanishing mass, $m|_{r=0}=0$, at the centre of the star, and a vanishing pressure, $p|_{r=R}=0$, at the surface ($r = R$). For the metric function $\nu(r)$, we start the integration from $r = 0$ with $\nu(0) = 0$ and the solution must match the exterior solution at the surface. We implement it by making the following variable change~\citep{GlendenningBook}
\begin{equation}\label{eq:nu}
    \nu(r) \longrightarrow \nu(r) - \nu(R) + \frac{1}{2}ln\left( 1 - \frac{2M}{R}\right),
\end{equation}
where $M = m(R)$ is the total mass of the star. \\

For the slowly rotating equilibrium configuration, in addition to the TOV equations, we need to solve an equation for the other metric function $\omega(r)$ given by the Hartle equation~\citep{Hartle}
\begin{equation}\label{eq:Hartle}
    \frac{1}{r^4}\frac{d}{dr}\left(r^4j\frac{d\Bar{\omega}}{dr}\right) + \frac{4}{r}\frac{dj}{dr}\Bar{\omega} = 0,
\end{equation}
where 
\begin{equation}\label{eq:rotation}
    \Bar{\omega}(r) = \Omega - \omega (r).
\end{equation}
j(r) is defined in terms of the metric functions as 
\begin{align}
    j(r) &= e^{-(\nu+\lambda)} \text{\hspace*{6mm} for $r\leq R$,} \nonumber\\
    &= 1 \text{\hspace*{16mm} for $r>R$.}
\end{align}
The differential equation~\eqref{eq:Hartle} for $\omega (r)$ can be integrated from r = 0 with an arbitrary choice of the  central value $\Bar{\omega}(0)$ and a vanishing slope~\citep{GlendenningBook}. At the surface, it should match the exterior solution. From equation~\eqref{eq:Hartle}, the exterior solution ($r>R$) is given by
\begin{equation}\label{omegaext}
    \omega(r) = \frac{2J}{r^3},
\end{equation}
where J is the angular momentum of the star. At the surface, the corresponding boundary conditions should be matched 
\begin{eqnarray}
J &=& \frac{1}{6}R^4\left(\frac{d\Bar{\omega}}{dr}\right)_R, \nonumber \\
\Omega &=& \Bar{\omega}(R) + \frac{2J}{R^3}. \nonumber \\
\label{eq:matchrot}
\end{eqnarray}
Since, $\Bar{\omega}(r)$ depends on the rotational frequency $\Omega$,  we normalise $\Bar{\omega}(r)$ by $\Omega$, $\Bar{\omega}(r) \equiv \frac{\Bar{\omega}(r)}{\Omega}$ and make $\Omega = 1$. \\

The tidal deformability parameter quantifies the degree of the tidal deformation effects due to the companion in coalescing binary NS systems during the early stages of an inspiral. It is defined as 
\begin{equation}\label{eq:tidal}
    \lambda = - \frac{Q_{ij}}{\varepsilon_{ij}},
\end{equation}
where $Q_{ij}$ is the induced mass quadrupole moment of the NS and $\varepsilon_{ij}$ is the gravitational tidal field of the companion. The dimensionless tidal deformability ($\Lambda$) can be obtained by solving a set of differential equations coupled with the TOV equations and it is related to the dimensionless $l = 2$ tidal Love number $k_2$~\citep{Hinderer,Hinderer2008} as
\begin{equation}
    \Lambda = \frac{2}{3} k_2 \left( \frac{R}{M} \right)^5.~
\label{eq:love}
\end{equation}

\subsection{Perturbation equations}
Here, we consider the non-radial perturbations of these slowly rotating equilibrium models to linear order in $\Omega$. Since the equilibrium spacetime is stationary and axisymmetric, we decompose our perturbations using the Lagrangian formalism into modes of
the form $e^{i(\sigma t+m\phi)}$~\citep{Lockitch2001}. We express the perturbed configuration in terms of the set $(h_{\alpha\beta},\zeta^{\alpha},\delta\varepsilon,\delta p)$. Since, the perturbed energy density and pressure are scalar, they have polar parity and given as 
\begin{equation}\label{eq:perener}
   \delta\varepsilon = \delta\varepsilon(r)Y_l^m, \delta p = \delta p(r)Y_l^m.
\end{equation}
The Lagrangian displacement vector is defined as 
\begin{equation}\label{eq:dispvector}
\begin{split}
      \zeta^{\alpha} &= \frac{1}{i\kappa\Omega}\sum_{l=m}^{\infty}\left(\frac{1}{r}W_l(r)Y_l^mr^{\alpha} + V_l(r)\nabla^{\alpha}Y_l^m  \right.\\ &  \left. \quad - iU_l(r)P^{\alpha}_{\nu}\epsilon^{\nu\beta\gamma\delta}\nabla_{\beta}Y^l_m\nabla_{\gamma}t\nabla_{\delta}r\right)e^{i\sigma t} ,
\end{split}
\end{equation}
where 
\begin{equation}
     P^{\alpha}_{\nu} = e^{\nu+\lambda}(\delta_{\nu}^{\alpha}-t_{\nu}\nabla^{\alpha}t),
\end{equation}
and the co-moving frequency is given by
\begin{equation}\label{kdef}
    \kappa\Omega \equiv \sigma + m\Omega .
\end{equation}
The perturbation variables $W_l \& V_l$ have polar parity and $U_l$ has axial parity. In the Regge-Wheeler gauge the metric perturbation is given by 
\begin{equation}\label{eq:pertmetric}
\begin{split}
      h_{\mu\nu} =Y_l^me^{i\sigma t} \sum_{l=m}^{\infty} \hspace*{6cm}&\\
    \begin{bmatrix}
     H_{0,l}(r)e^{2\nu} & H_{1,l}(r) & h_{0,l}(r)(\frac{m}{\sin(\theta)}) & ih_{0,l}(r)\sin(\theta) \partial_{\theta}\\
     H_{1,l}(r)e^{2\nu} & H_{2,l}(r) & h_{1,l}(r)(\frac{m}{\sin(\theta)}) & ih_{0,l}(r)\sin(\theta) \partial_{\theta}\\
     symm & symm & r^2K_l(r) & 0 \\
     symm & symm & 0 & r^2\sin^2(\theta)K_l(r)
    \end{bmatrix},
\end{split}
\end{equation}
which contains both axial ($h_{0,l},h_{1,l}$) and polar ($H_{0,l},H_{1,l},H_{2,l},K_l$) parity components. We obtain the perturbation equations by requiring $\delta G_{\alpha\beta} = 8\pi \delta T_{\alpha\beta}$ term by term. Since we are considering upto order $\Omega$ variables and the displacement vector in Eqn.~\eqref{eq:dispvector} has already a $\kappa\Omega$ term, we only keep the zeroth order perturbation variables. The variables can be grouped depending upon their order of dependence on the rotational frequency $\Omega$. The relevant variables which are zeroth order in $\Omega$ are $W_l, V_l, U_l, H_{1,l}, h_{0,l}$. Because $h_{1,l}$ is an order $\Omega$ variable, we drop the ``0’’ subscript and write  $h_{0,l}$ as $h_l$. The relevant $O(1)$ equations are given by    \\
\begin{equation}\label{eq:perturb0}
    H_{1,l} + \frac{16\pi(\varepsilon + p)}{l(l+1)}e^{2\lambda}rW_l = 0~,
\end{equation}
\begin{equation}\label{eq:perturb1}
    V_l[l(l+1)(\epsilon + p)] - e^{-(\nu+\lambda)}[(\epsilon + p)e^{\nu+\lambda}W_l]' = 0,
\end{equation}
\begin{equation}\label{eq:perturb2}
\begin{split}
      r^2h_l'' - r^2(\nu'+\lambda')h_l' & + [(2-l^2-l)e^{2\lambda} \\- r(\nu'+\lambda') -2]h_l  &- 4r(\nu'+\lambda')U_l = 0,
\end{split}
\end{equation}
where a prime denotes a derivative with respect to r. We have used Eq.~\eqref{eq:perturb0} to eliminate the variable $H_{1,l}$ in favour of $W_l$. \\

 To close the system of equations we obtain two other independent equations that arise at $O(\Omega)$ which enforces the conservation of vorticity in constant entropy surfaces
\begin{align}\label{eq:perturb3}
   & 0 = [l(l+1)\kappa\Omega(h_l+U_l) - 2m\Bar{\omega}U_l] \nonumber \\
   & + (l+1)Q_l[(2\Bar{\omega}+\mu)W_{l-1} - 2(l-1)\Bar{\omega}V_{l-1}] \nonumber \\
   & - Q_{l+1}[(2\Bar{\omega}+\mu)W_{l+1} + 2(l+2)\Bar{\omega}V_{l+1}],
\end{align}
\begin{equation}\label{eq:perturb4}
\begin{split}
  0 &= (l-2)l(l+1)Q_{l-1}Q_l\left[-2\Bar{\omega}rU'_{l-2}   \right. \\ & \left. + 2(l-1)\Bar{\omega}U_{l-2} + (l-3)\mu U_{l-2}\right] \\
  & +(l+1)Q_l\left[(l-1)l\kappa\Omega rV'_{l-1} - 2ml\Bar{\omega}rV'_{l-1} \right. \\ & \left. -2l(l-1)\kappa\Omega r\nu'V_{l-1}+2ml(l-1)\Bar{\omega}V_{l-1}  \right. \\ & \left.  + m(l-3)l\mu V_{l-1}\right] -(l+1)Q_l \nonumber\\
  & \left[(l-1)l\kappa\Omega e^{2\lambda}W_{l-1} - 4\kappa\Omega r(\nu'+\lambda')W_{l-1} \right] \nonumber\\
  & +l(l+1)\left[m\kappa\Omega r(h'_l+U'_l) - 2m\kappa\Omega r\nu'(h_l+U_l) \right. \\ & \left. +((l+1)Q^2_l-lQ^2_{l+1})(2\Bar{\omega}rU'_l+2\mu U_l)\right] \nonumber\\
  & + l(l+1)\left[m^2+l(l+1)(Q^2_{l+1}+Q^2_l-1)\right] \nonumber\\
  & \times(2\Bar{\omega}+\mu)U_l-lQ_{l+1}\left[(l+1)(l+2)\kappa\Omega rV'_{l+1} \right. \\ & \left. +2m(l+1)\Bar{\omega}rV'_{l+1} - 2(l+1)(l+2)\kappa\Omega r\nu'V_{l+1}\right] \nonumber\\
  & -lQ_{l+1}\left[2m(l+1)(l+2)\Bar{\omega}V_{l+1} \right. \\ & \left.  + m(l+1)(l+4)\mu V_{l+1} -(l+1)(l+2)\kappa\Omega e^{2\lambda}W_{l+1} \right. \\ & \left. + 4\kappa\Omega r(\nu'+\lambda')W_{l+1} \right] \nonumber\\
  & + l(l+1)(l+3)Q_{l+1}Q_{l+2}\left[2\Bar{\omega}rU'_{l+2} \right. \\ & \left. + 2(l+2)\Bar{\omega}U_{l+2} + (l+4)\mu U_{l+2}\right],
\end{split}  
\end{equation}

where we have defined $\mu = re^{2\nu}(\Bar{\omega}e^{-2\nu})'$. Please note that the equation~\eqref{eq:perturb4} is a simplified form of Eq. (48) from ~\cite{idrisy2015}(or Eq. (22) of ~\cite{Lockitch2003}) which we have taken from Eq. (346) of ~\cite{lockitchthesis}. For the barotropic stars (which is the case for cold neutron stars), the conservation of vorticity  gives rise to a mixing of axial and polar modes at zeroth order in $\Omega$~\citep{Lockitch2001}.This suggests that the modes of barotropic stars will generically be of a hybrid nature.

\subsection{Boundary Conditions }
In order to solve the equations ~\eqref{eq:perturb1}~\eqref{eq:perturb2}~\eqref{eq:perturb3}~\eqref{eq:perturb4}, we need to apply the appropriate boundary conditions. The first boundary condition is also called the regularity condition which says that the perturbation equations must be regular at the centre of the star.  To implement this boundary condition, we introduce a new variable $\Tilde{F_l}$ corresponding to each perturbation variable $F_l$ as
\begin{equation}\label{eq:boundary1}
    F_l(r) = \left(\frac{r}{R}\right)^{l+q}\Tilde{F_l}(r),
\end{equation}
where $F_l$ is any one of the perturbation variables $W_l,V_l,U_l,h_l$. The axial parity variables ($U_l,h_l$) have $q=1$ and polar parity variables ($W_l,V_l$) have $q=0$~\citep{Lockitch2003}. \\

The next boundary condition comes from the fact that the Lagrangian perturbation of the pressure is zero at the surface of the star which translates to
\begin{equation}\label{eq:boundary2}
    W_l(R) = 0~.
\end{equation}
Now $h_l$ is the only variable defined outside the star also where it follows
\begin{equation}\label{eq:hl_evolve}
    \left(1-\frac{2M}{r}\right)\frac{d^2h_l}{dr^2} - \left[\frac{l(l+1)}{r^2} - \frac{4M}{r^3}\right]h_l = 0.
\end{equation}
The solution of this equation is given by the hypergeometric function $h_l(r) = _2F_1(l-1,l+2;2l+2;2M/r)$~\citep{idrisy2015}. The interior and exterior solutions are matched via the following boundary conditions 
\begin{equation}\label{eq:boundary3}
    \lim_{\epsilon \to 0}[h_l(R-\epsilon)-h_l(R+\epsilon)] = 0,
\end{equation}
\begin{equation}\label{eq:boundary4}
     \lim_{\epsilon \to 0}[h_l(R-\epsilon)h'_l(R+\epsilon)-h'_l(R-\epsilon)h_l(R+\epsilon)] =0.
\end{equation}

\section{Numerical Method}
\label{sec:numerical}
To get the r mode frequencies, we need to solve the perturbation equation ~\eqref{eq:perturb1}~\eqref{eq:perturb2}~\eqref{eq:perturb3}~\eqref{eq:perturb4} along with the boundary conditions ~\eqref{eq:boundary1}~\eqref{eq:boundary2}~\eqref{eq:boundary3}~\eqref{eq:boundary4}. For a given EOS, solving the TOV equations~\eqref{eq:tov} and the Hartle equation ~\eqref{eq:Hartle} gives the necessary equilibrium variables $p(r),\varepsilon(r),\nu(r),\lambda(r)$ and $\omega(r)$. Since the perturbation equations are coupled in terms of $l$, we need to set a upper limit to our $l$ value ($l_{max}$) up to which we will solve these equations. As we are focusing on axial-led hybrid modes, we fix $l_{max}$ to an odd value to get a closed system of equations~\citep{idrisy2015}. For this choice of axial-led hybrid modes, the perturbation variables has contribution only from the terms with~\citep{Lockitch2001} 
\begin{eqnarray*}\label{eq:contri}
 \text{axial parity($h_l,U_l$) with } l &=& m, m+2, m+4, \nonumber\\
 \text{polar parity($W_l,V_l$) with } l &=& m+1, m+3, m+5~. \nonumber\\
\end{eqnarray*}
So, we solve for the eigenfunctions $h_l,U_l,W_{l+1},V_{l+1}$ for $l = m,m+2,...$ and set others to zero. \\
Instead of integrating the coupled perturbation equations, we adopt a spectral method using the Chebyshev polynomials to find the eigenfrequency $\kappa$ similar to~\cite{Lockitch2003,idrisy2015}. We express our system of ordinary differential equations in terms of sum of basis functions, Chebyshev polynomials in our case and using some useful identities, we reduce these differential equations in a system of algebric equations which is then solved using root-finding techniques to get the eigenfrequency. 
\subsection{Chebyshev Polynomials}
The Chebyshev polynomials of first kind are defined in $[-1,1]$ range by
\begin{equation}\label{eq:chebyshevdef}
    T_i(y) = \cos(i \arccos (y))~.
\end{equation}
Any function S(y) expanded in terms Chebyshev polynomials are given by 
\begin{equation}\label{eq:chevyshev_exp}
    S(y) = \sum_{i=0}^{i_{max}}s_i T_i(y) - \frac{1}{2}s_0,
\end{equation}
where 
\begin{equation}\label{eq:chebyshevdef_coff}
\begin{split}
    s_i &= \frac{2}{i_{max}}\sum_{j=0}^{i_{max}+1} \left[S\left[\cos\left(\frac{\pi(j+\frac{1}{2})}{i_{max}}\right)\right] \right.\\ & \left. \cos\left(\frac{\pi i(j+\frac{1}{2})}{i_{max}}\right)\right].
\end{split}    
\end{equation}
Since our variable r is in domain [0,R], we define a new variable $y = 2\left(\frac{r}{R}\right)-1$. Now, choosing  $i_{max}$ is a trade-off as increasing $i_{max}$ gives better convergence but it also increases the number of equations to solve. Along with this, we also make use of two other identities for Chebyshev polynomials involving derivatives of a function and product of functions.  \\

If $f_l'$ and $f_l$ are the Chebyshev coefficients of the derivative of a function and the function itself respectively, then they are related by 
\begin{equation}\label{eq:chebyshev_deriv}
    f'_{l,i} - f'_{l,i+2} = 2(i+1)f_{l,i+1}.
\end{equation}
If $b_i$ and $f_{l,i}$ are the Chebyshev coefficients of a background function B(y) and perturbation variable $F_l(y)$ respectively, then the Chebyshev coefficients for their products are given by
\begin{equation}\label{eq:chebyshev_product}
    \pi_{l,i} =  \sum_{j=0}^{i_{max}}[b_{i+j}+\Theta(j-1)b_{|i-j|}]f_{l,j},
\end{equation}
where $\Theta$ is the step function. \\

After imposing the regularity condition~\eqref{eq:boundary1},each term in the perturbation equation can be written in terms of a background function($B(r)$) which depends on the star equilibrium profiles and a foreground function($\Tilde{F_l}(r)$) which are the perturbation variables. Since, $f'_{l} $ and $f_{l}$ are connected by the Eq.~\eqref{eq:chebyshev_deriv}, we only have $f_{l}$ as our unknown functions. Now we expand each of them in terms of the chebyshev polynomials and simplify them using the identities~\eqref{eq:chebyshevdef_coff}~\eqref{eq:chebyshev_deriv}~\eqref{eq:chebyshev_product}. We also re-write the boundary conditions in the same way. \\

After expanding all the perturbation equations in terms of Chebyshev polynomials, we extract the co-efficients using the identities~\eqref{eq:chebyshevdef_coff}~\eqref{eq:chebyshev_deriv}~\eqref{eq:chebyshev_product}. This leads to a system of $2(l_{max}-3)i_{max}$ linear equations for $\kappa$ with the unknown functions $f_{l}$. We represent the  system of equations as 
\begin{equation}
    A(\kappa)x = 0,
\end{equation}
where x is the vector of ${f_l}$s containing Chebyshev coefficients of the perturbation variables. To incorporate the boundary conditions, we replace the equation that came from the highest order Chebyshev coefficient $f_{i_{max}}$ by a boundary condition~\citep{idrisy2015}. For example, to implement the boundary condition $W_l(R) = 0$~\eqref{eq:boundary2}, we replace the highest order Chebyshev coefficient $f_{i_{max}}$ as 
\begin{equation}
    w_{l_{i_{max}}} = \frac{1}{2}w_{l_0} - \sum_{i=0}^{i_{max}-1}w_{l_i},
\end{equation}
where $w_l$'s are the Chebyshev coefficients of the variable $W_l$. We have used at at $r = R$, $y = 1$ and $T_i(1) = 1$  $\forall i$.

\subsection{Root finding method}
To find the eigenfrequency $\kappa$, we set $det(A(\kappa)) = 0$. This leads to a very high degree of polynomial in $\kappa$ for any reasonable value of $l_{max}$  and $i_{max}$ which is not solvable by any standard root-finding techniques. We use the second root-finding algorithm given in ~\cite{idrisy2015} which uses the Singular Value Decomposition (SVD) of the matrix A, SVD(A) =  {$U\Sigma V$}. We vary $\kappa$ in the physically possible range ($0.67$-$0.4$) and look for the value of $\kappa$ that results in the smallest value for the last element on the diagonal of $\Sigma$. For a particular $l_{max}$ and $i_{max}$, we find several roots for $\kappa$. The way to determine the correct root is to start at small values of $l_{max}$ and $i_{max}$ and increase them step by step. We will always converge to the correct root for any $l_{max}$ and $i_{max}$ while the others will change unpredictably. \\
In Fig. 1 we plot the last diagonal element in SVD of the matrix A vs $\kappa$ for different choice of $i_{max}$ with fixed $l_{max}$ and vice versa for an $n = 1$ polytrope with compactness of 0.153. 
\begin{figure}[ht]
\centering
\begin{minipage}[b]{.5\textwidth}
  \centering
  \includegraphics[width=\linewidth]{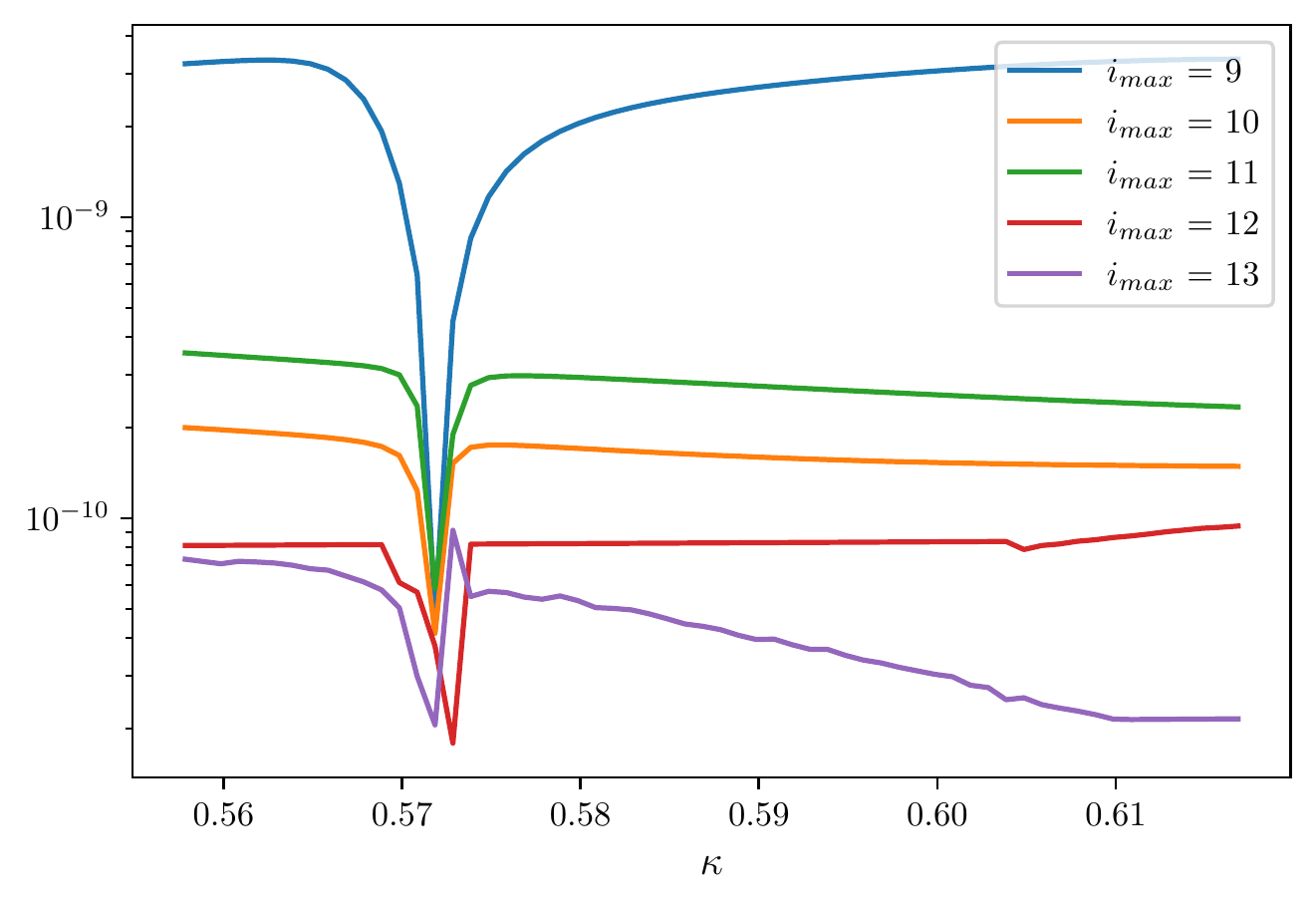}
  \label{fig:sub1}
\end{minipage}%
\\
\begin{minipage}[b]{.5\textwidth}
  \centering
  \includegraphics[width=\linewidth]{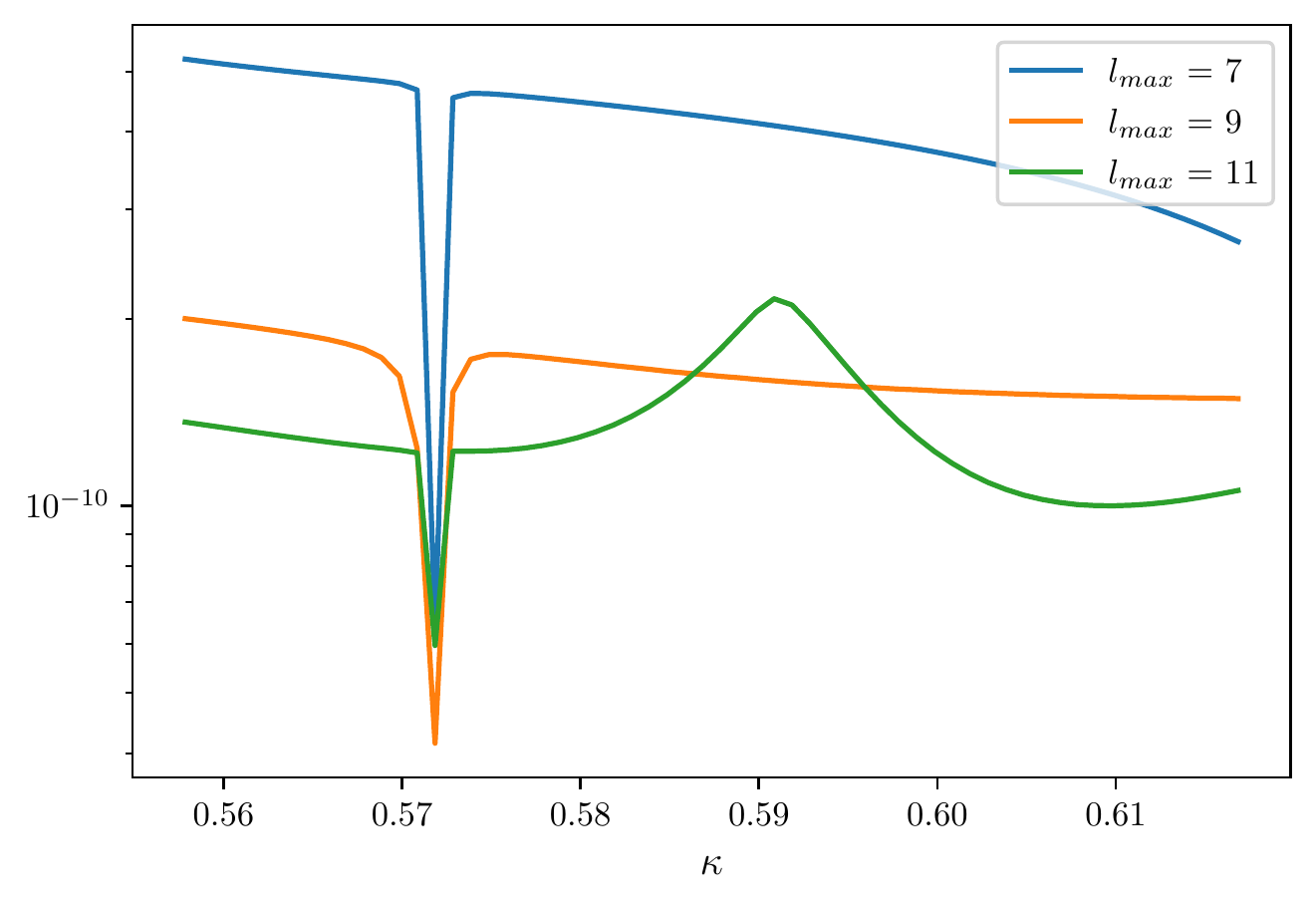}
  \label{fig:sub2}
\end{minipage}
\caption{Convergence of the root with changing $l_{max}$ and $i_{max}$. The upper panel shows the convergence for varying $i_{max}$ while $l_{max} = 9$ and the lower panel shows convergence for varying $l_{max}$ for $i_{max} = 10$}
\label{fig:test}
\end{figure}
For the polytropes we get convergence at minimum of $l_{max} = 7$ and $i_{max} = 9$ and the convergence stops above  $l_{max} = 13$ and $i_{max} = 13$. For both polytropes and realistic equation of states, we get convergence upto 3 orders after decimal place. Sometimes, the converged root will vary in the 3rd decimal place with different $l_{max}$ and $i_{max}$. We take the statistical mode of the roots for different combinations of $l_{max}$ and $i_{max}$.

\section{Results }
\label{sec:results}
\subsection{Polytropic EOS}
\begin{figure}[ht]
\centering
\includegraphics[width=1.\linewidth]{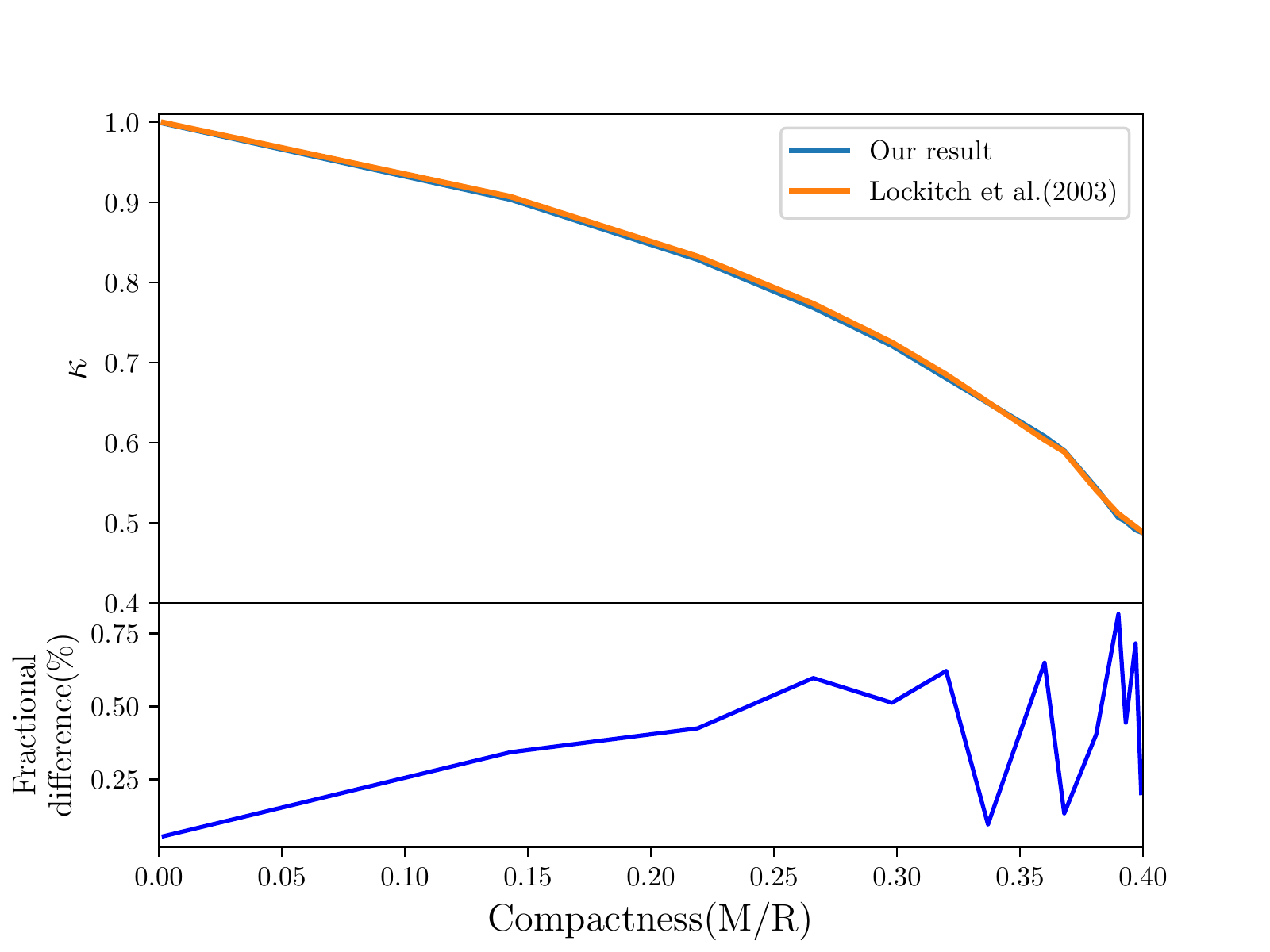}
\caption{Normalised eigenfrequency $\kappa/\kappa_N$ of $m = 2$ mode with their Newtonian counterparts for uniform density star as a function of compactness. Relative differences are shown in the bottom panel.}
\label{fig:poltr_lockitch}
\end{figure}

\begin{figure}[ht]
\centering
\includegraphics[width=1.\linewidth]{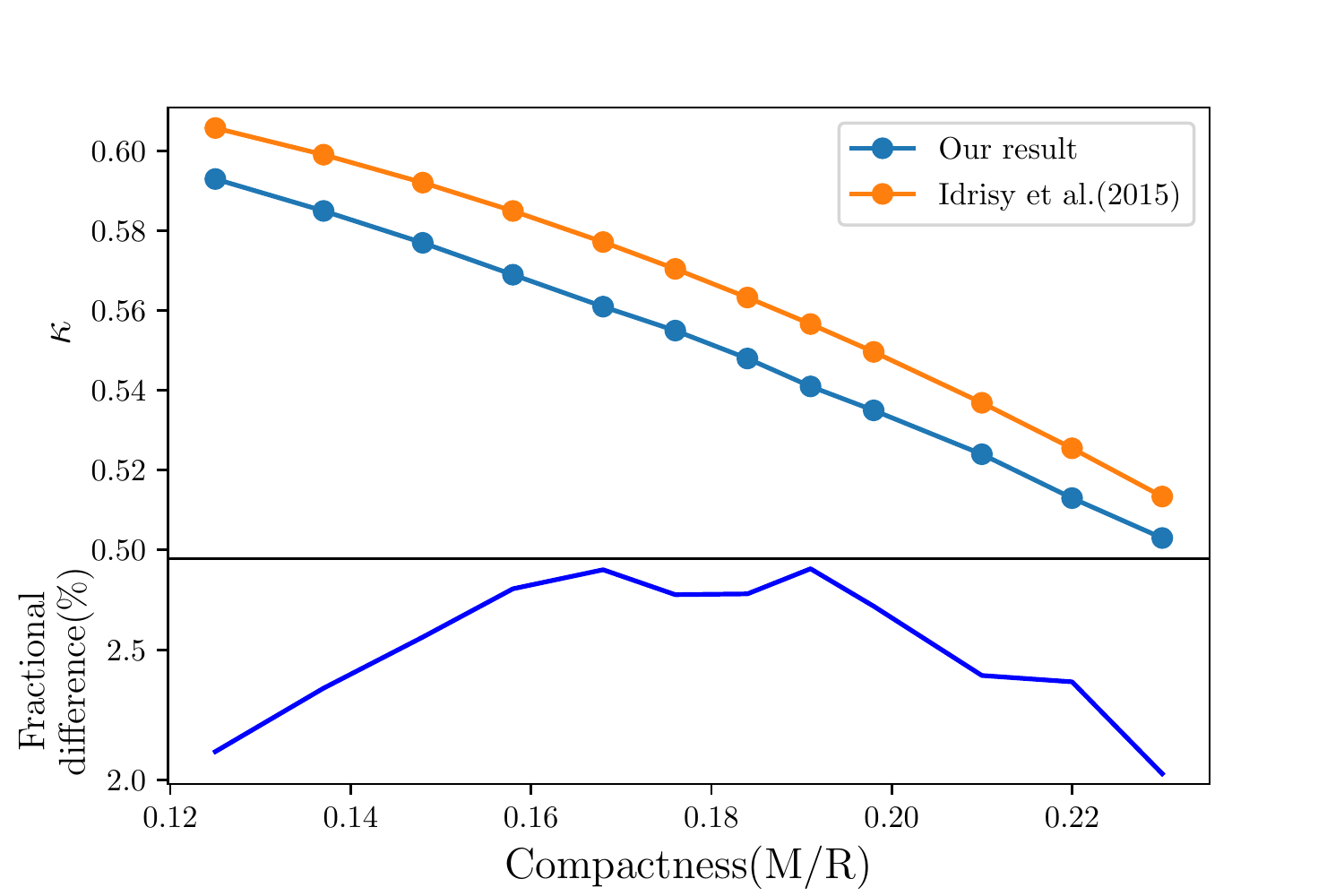}
\caption{Eigenvalue $\kappa$ for equilibrium stars  of $n = 1$ polytrope for different compactness. Relative differences are shown in the bottom panel.}
\label{fig:poltr_idrisy}
\end{figure}

For the Newtonian stars, the $l = m = 2$ r-mode is the one expected to dominate the gravitational wave radiation from the hot and fast rotating Neutron stars~\citep{Andersson1999,Lindblom1998}. But, ~\cite{Lockitch2001} showed that for the relativistic barotropes pure $l = m = 2$ does not exist. The corresponding modes are axial-led hybrid modes with $m = 2$.  To test the accuracy of our numerical code, we first consider the case for polytropic EOS and  compare our results with previous studies~\citep{idrisy2015,Lockitch2003}. In Fig~\ref{fig:poltr_lockitch}, we plot the r-mode frequency as a function of compactness for uniform density star model or $n = 0$ polytropic EOS and compare the same with Fig (1) in ~\cite{Lockitch2003}. We get a significant match and the relative error is around $0.1-0.7\%$. In Fig~\ref{fig:poltr_idrisy}, we plot the r-mode frequency as a function of compactness for $n = 1$ polytropic EOS along with the quadratic fit relation between r-mode frequency and compactness given by Eq. (69) in ~\cite{idrisy2015}. In this case, we see a greater deviation (around $2-3\%$) in the r-mode frequency from the results in \cite{idrisy2015}. We discuss the implications and possible reasons for this difference in Sec.~\ref{sec:discussion}\\

\begin{figure}[ht]
\centering
\includegraphics[width=\linewidth]{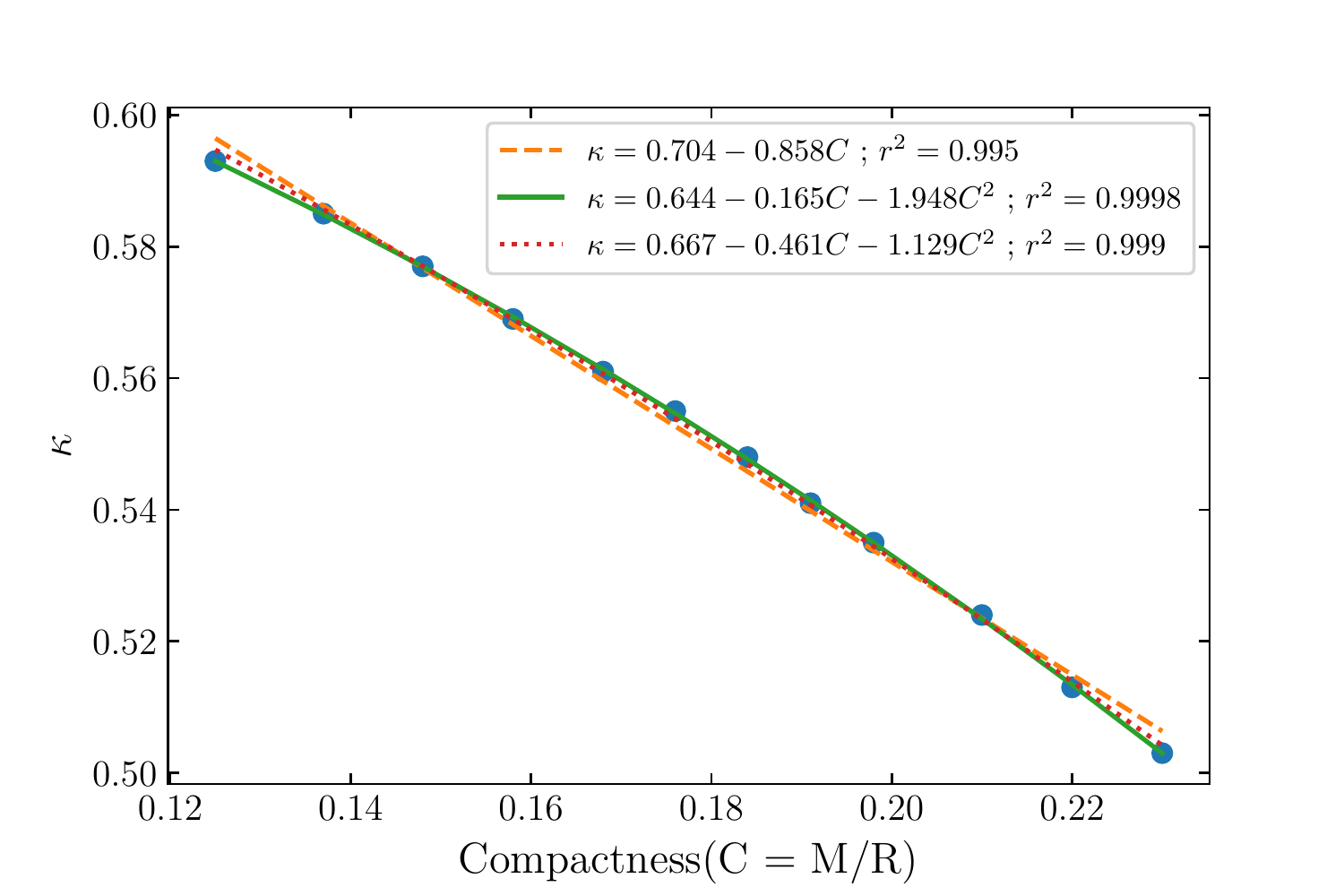}
\caption{Linear and Quadratic fits to $\kappa$ vs compactness results for equilibrium stars  of $n = 1$ polytrope}
\label{fig:poltr_fit}
\end{figure}

In Fig~\ref{fig:poltr_fit}, we show the best linear and quadratic model fits using Least Squares method to our results. For quadratic models, we also use a second model where the zeroth order term of the polynomial is fixed to $2/3$ to constrain the fact that as $M/R\longrightarrow 0$, we reach Newtonian limit where the r-mode frequency for $l=m=2$ mode is given by $$\kappa_N = \frac{2}{m+1} = 2/3~.$$ \\
From the $R^2$ value of the fits, we see that any quadratic model is a better fit than the linear model. So, we use the quadratic model which satisfies the Newtonian limit
\begin{equation}\label{eq:univ_poltr}
    \kappa = 0.667 - 0.461C - 1.129C^2.
\end{equation}
We also see that along with the negative quadratic term, we also have the linear term negative in our fit equation~\eqref{eq:univ_poltr}. This also satisfies the physical constraint that the r-mode frequency should decrease as we increase compactness~\citep{Lockitch2001} unlike the Eq. (69) in ~\cite{idrisy2015} which implies that $\kappa$ increases with increasing compactness for $M/R < 0.10$ due to the positive linear term. 

\subsection{Realistic equations of state }
\subsubsection{Tabulated equation of state}
For realistic equation of states, ~\citep{idrisy2015} considered 14 EOSs under the constraint that the EOS could support a minimum of $1.85M_{\odot}$. From the recent observations, the heaviest known pulsar PSR J0740+6620 has a maximum mass of $2.08^{+0.07}_{-0.07}$ $M_{\odot}$~\citep{fonseca2021refined}. Taking a upper limit of 1-$\sigma$ confidence interval of this maximum observed pulsar mass, we only consider EOS that support a minimum of $2.01M_{\odot}$ neutron star which rules out 3 EOSs - GNH3,BBB2 and ALF4 used in ~\cite{idrisy2015}. Also, the  recent analyses of the GW170817 event~\citep{Abbott2019} apply a constraint on the upper bound of the effective tidal deformability $\tilde{\Lambda} <$ 720~\citep{Tong2020} using the  PhenomPNRTwaveform model and low-spin highest posterior density interval for tidal deformability. Using these multi-messenger observations of neutron stars, 2 very stiff EOSs (MS1,MS1b) considered in ~\cite{idrisy2015} have been ruled out with good confidence~\citep{Abbott_2020,Biswas2022}. Along with the 9 remaining EOSs from~\cite{idrisy2015}, we consider 6 additional EOSs that satisfy the multi-messenger constraints. Out of these additional 6 EOSs, one model QHC19~\citep{Baym2019} incorporates a transition between a hadronic phase in the crust and a quark matter phase in the core and one other EOS model CMF5~\citep{Dexheimer_2008} includes 
nucleons and hyperons.  The other equation of state models describe purely nucleonic matter.  All our EOS tables are obtained from either COMPOSE ~\citep{Compose,Oertel2016} or an EOS catalogue from~\cite{lalsim} used in LALSuite~\citep{lalsuite}.  For the EOS tables obtained from COMPOSE, we use their  first-order interpolation option~\citep{Oertel2016} and for the tables from~\cite{lalsim} we use standard cubic spline interpolation~\citep{Abbott_2020}. All the EOSs with their maximum mass and radius at $1.4M_{\odot}$ are listed in Table~\ref{tab:eos}. \\

\begin{figure}[ht]
\centering
\includegraphics[width=\linewidth]{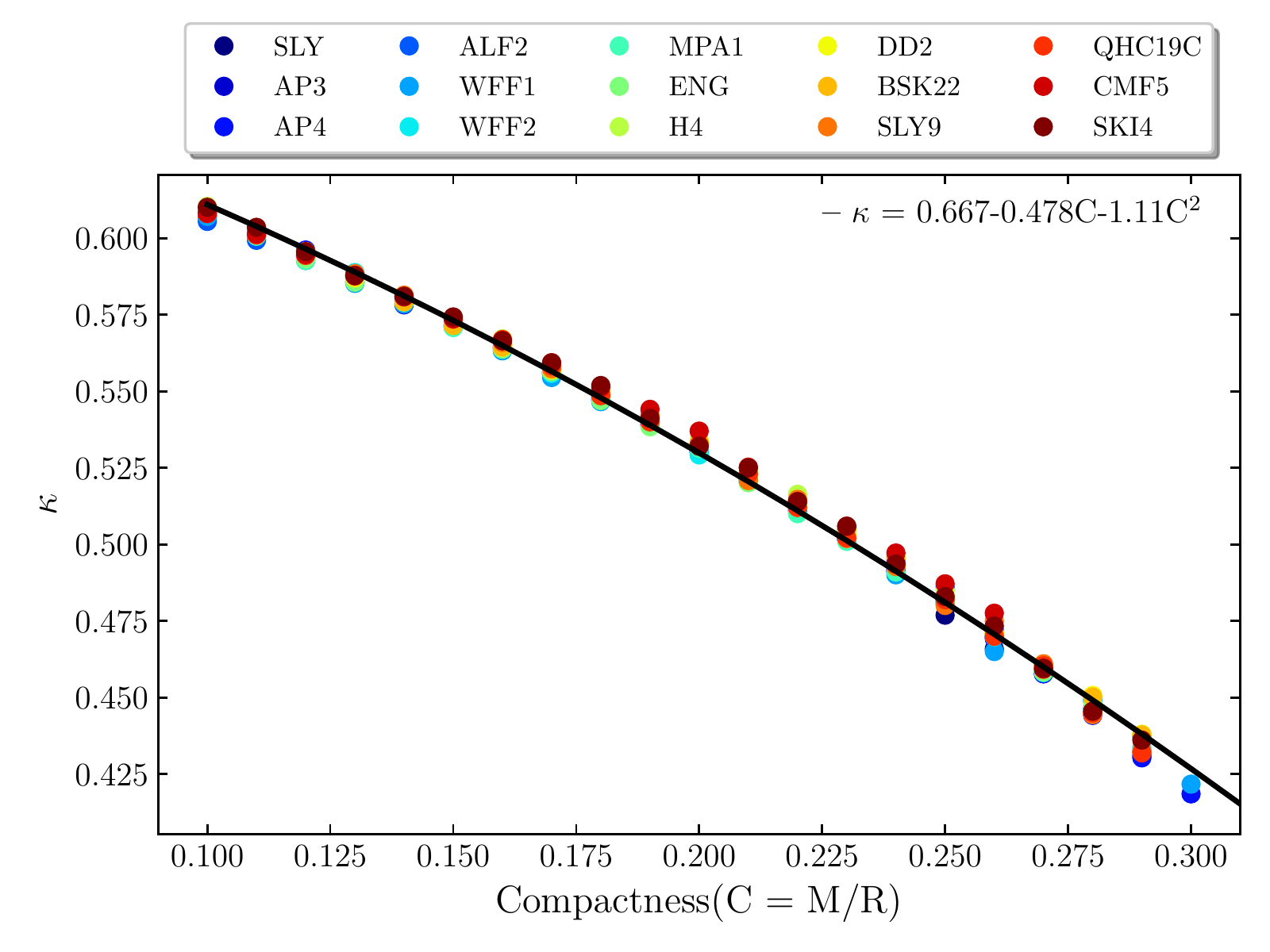}
\caption{ $\kappa$ vs compactness results for all 15 EOSs. The black line represents the best fir to our quadratic model.}
\label{fig:f_tabular}
\end{figure}

We compute the r-mode frequency $\kappa$ for these EOS models for a range of compactness varying from 0.1 to a maximum of 0.31. This compactness spans the range of possible neutron stars. In Fig~\ref{fig:f_tabular}, we plot the r-mode frequency $\kappa$ vs compactness for all the 15 EOSs and see that $\kappa$ does not change much as a function of compactness for different EOSs. 
We use a quadratic fit with the zeroth term fixed to its Newtonian value $2/3$ and get the universal relation for $\kappa$ vs compactness C as 
\begin{equation}\label{univ_real}
    \kappa = 0.667 - 0.478C - 1.11C^2.
\end{equation}
 
\begin{figure}[ht]
\centering
\includegraphics[width=\linewidth]{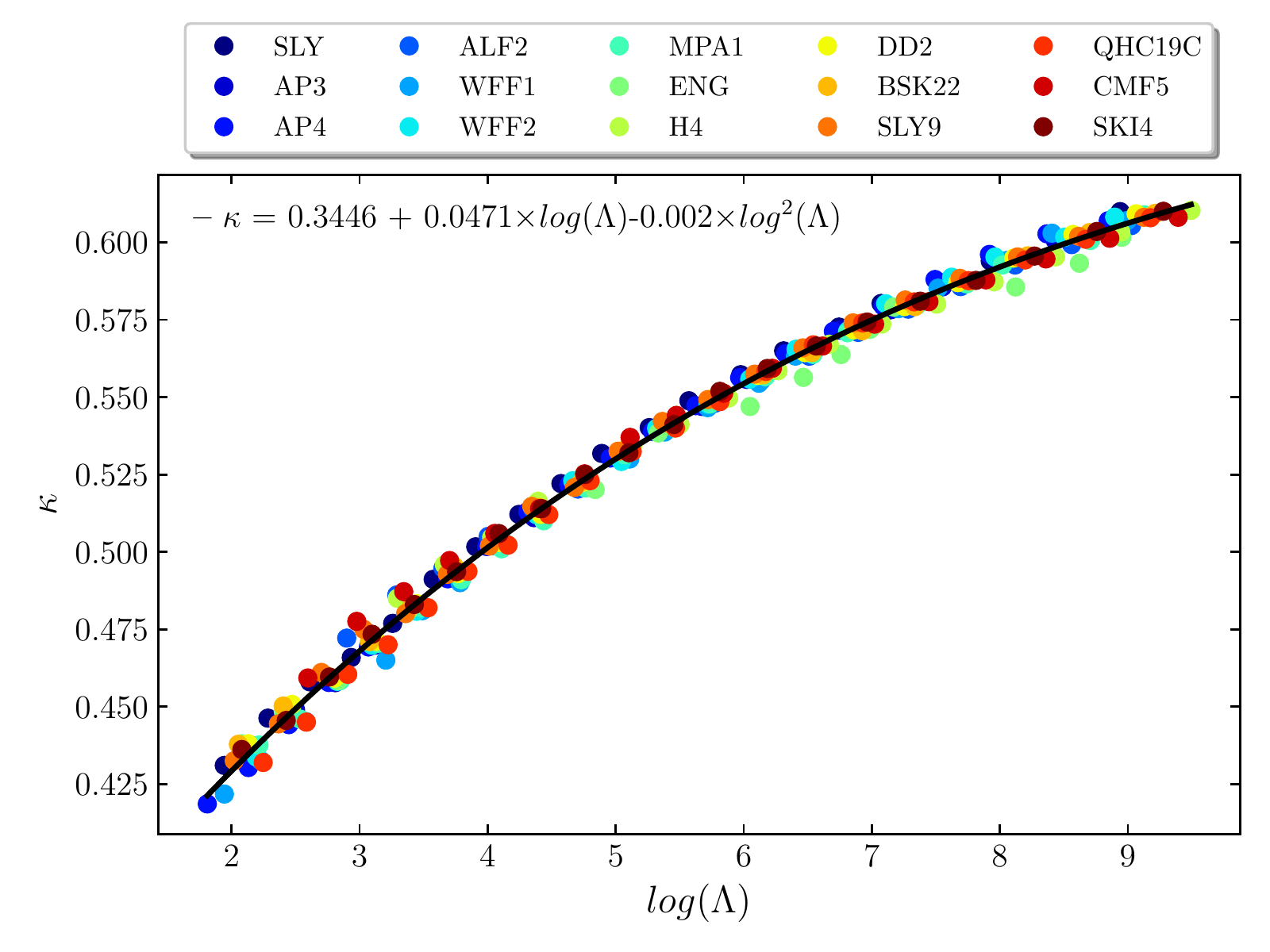}
\caption{ $\kappa$ vs $log(\Lambda)$ results for all 15 EOSs. The black line represents the best fir to our quadratic model.}
\label{fig:lamb_tabular}
\end{figure}

 To compare the results from~\cite{pawan2022,Ma2021} where they looked for possible resonant r-modes detection in the inspiral phase of the binary mergers using the third generation detector Einstein telescope, in Fig~\ref{fig:lamb_tabular}, we plot $\kappa$ as a function of $log(\Lambda)$ for all the 15 EOSs and fit it to a quadratic model.
The corresponding universal relation is given by
 \begin{equation}\label{eq:univ_lamb}
     \kappa = 0.3446 +0.0471 \> log(\Lambda) - 0.002 \> log^2(\Lambda).
 \end{equation}
 Comparing the result from~\cite{pawan2022}, we get around $5$-$6\%$ difference in the r-mode frequency for a given tidal deformability. In Table~\ref{tab:eos}, we also report the value of r-mode frequency $\kappa$ for a compactness of 0.15 and a tidal deformability of 400 for each of the 15 EOSs.

\begin{table}[ht]
\caption{List of all the tabulated EOSs with corresponding maximum mass, radius for $1.4M_{\odot}$(in km) star, $\kappa$ for a compactness of 0.15 (along with the same from ~\cite{idrisy2015}) and tidal deformability of 400.}
\centering
\begin{tabular}{||c |c |c| c| c||} 
 \hline
EOS & $M_{max}$ & $R_{1.4M_{\odot}}$ & $\kappa_{.15C}$(~\cite{idrisy2015})& $\kappa_{400\Lambda}$ \\ [0.5ex] 
 \hline\hline
 SLY & 2.05 & 11.77 & 0.573(0.587) & 0.558  \\
 \hline
 AP3 & 2.39 & 12.06 & 0.572(0.588) & 0.555 \\
 \hline
 AP4 & 2.21 & 11.40 & 0.571(0.587) & 0.557 \\
 \hline
 ALF2 & 2.09 & 13.18 & 0.571(0.588) & 0.552 \\
 \hline
 WFF1 & 2.13 & 10.40 & 0.572(0.587) & 0.551 \\
 \hline
 WFF2 & 2.20 & 11.14 & 0.571(0.587) & 0.554 \\
 \hline 
 MPA1 & 2.46 & 12.48 & 0.571(0.588) & 0.553 \\
 \hline
 ENG & 2.25 & 12.08 & 0.572(0.588) & 0.546 \\
 \hline
 H4 & 2.03 & 12.95 & 0.573(0.591) & 0.552 \\
 \hline
 DD2 & 2.42 & 13.04 & 0.572(-) & 0.554 \\
 \hline
 BSK22 & 2.26 & 12.73 & 0.571(-) & 0.554 \\
 \hline
 SLY9 & 2.16 & 12.20 & 0.574(-) & 0.555 \\
 \hline
 QHC19 & 2.18 & 11.35 & 0.574(-) & 0.553 \\
 \hline
 CMF5 & 2.07 & 12.87 & 0.573(-) & 0.554 \\
 \hline
 SKI4 & 2.17 & 12.10 & 0.574(-) & 0.556 \\
 \hline
\end{tabular}
\label{tab:eos}
\end{table}

\subsubsection{Nonparametric EOS model}
\begin{figure}[htbp]
\centering
\includegraphics[width=\linewidth]{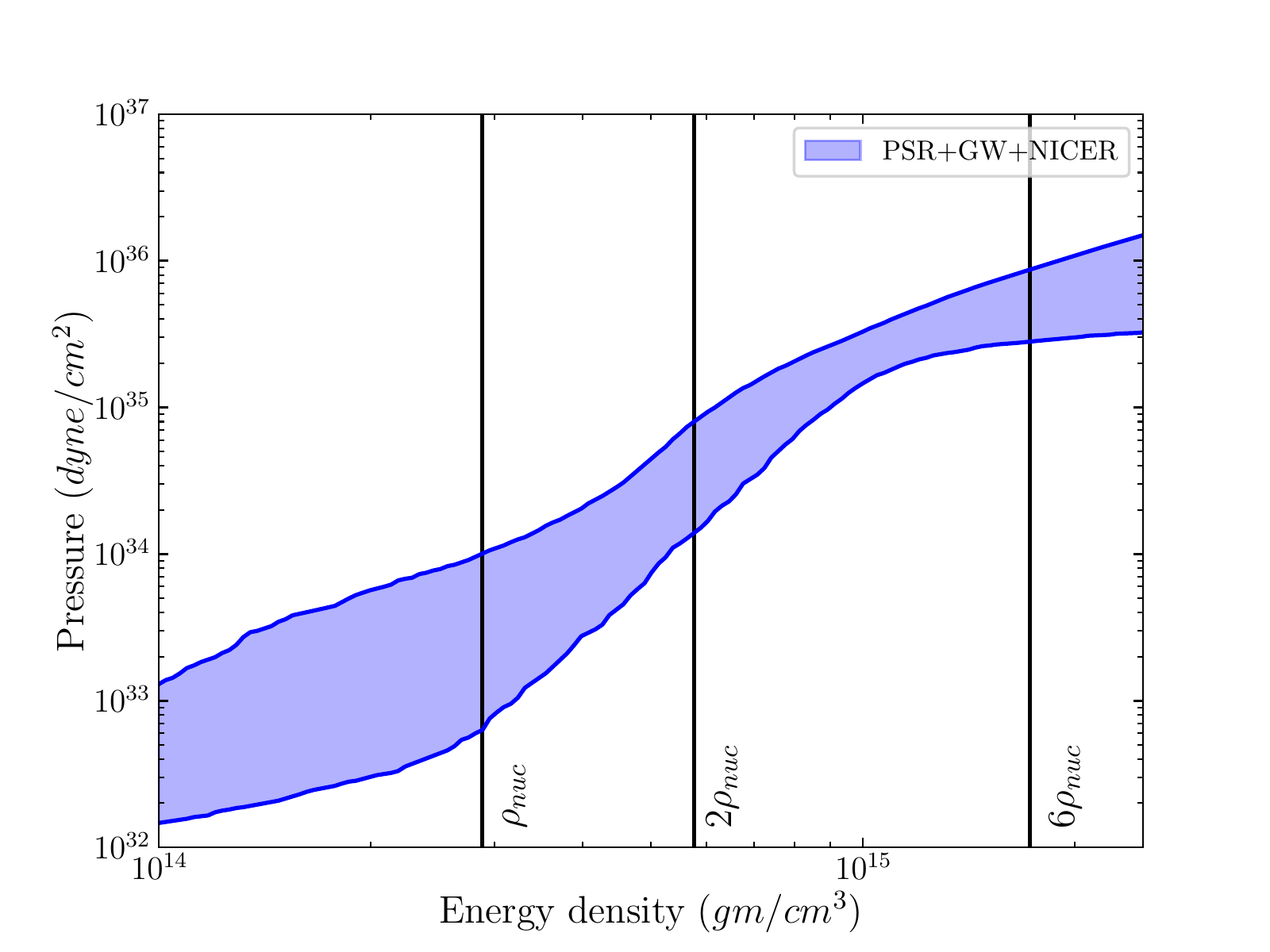}
\caption{95$\%$ confidence interval for pressure-density relation with constraints from GW, NICER and PSR mass measurement}
\label{fig:phil_eos}
\end{figure}

\begin{figure}[ht]
\centering
\begin{minipage}{.5\textwidth}
  \centering
  \includegraphics[width=\linewidth]{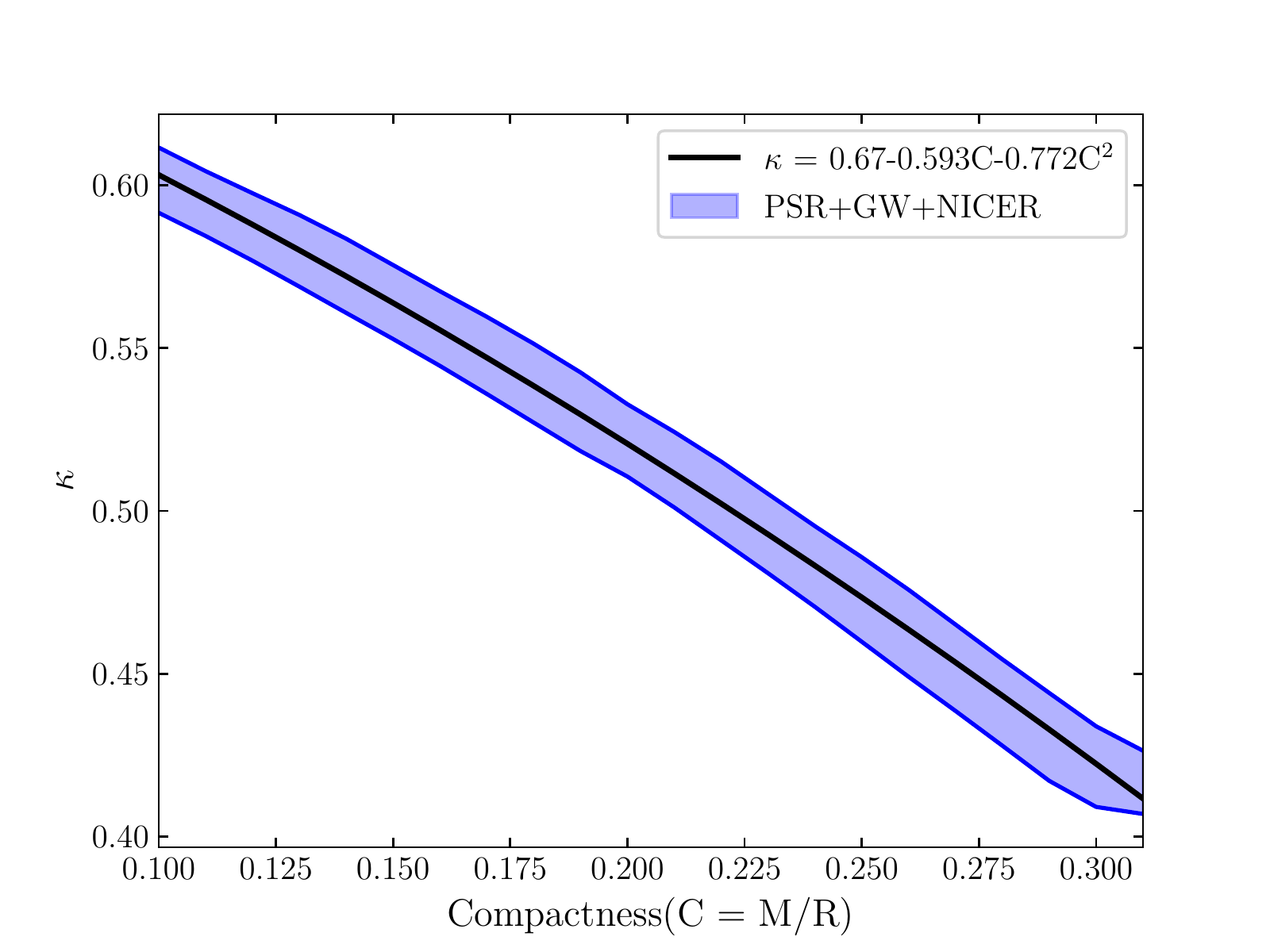}
\end{minipage}%
\\
\begin{minipage}{.5\textwidth}
  \centering
  \includegraphics[width=\linewidth]{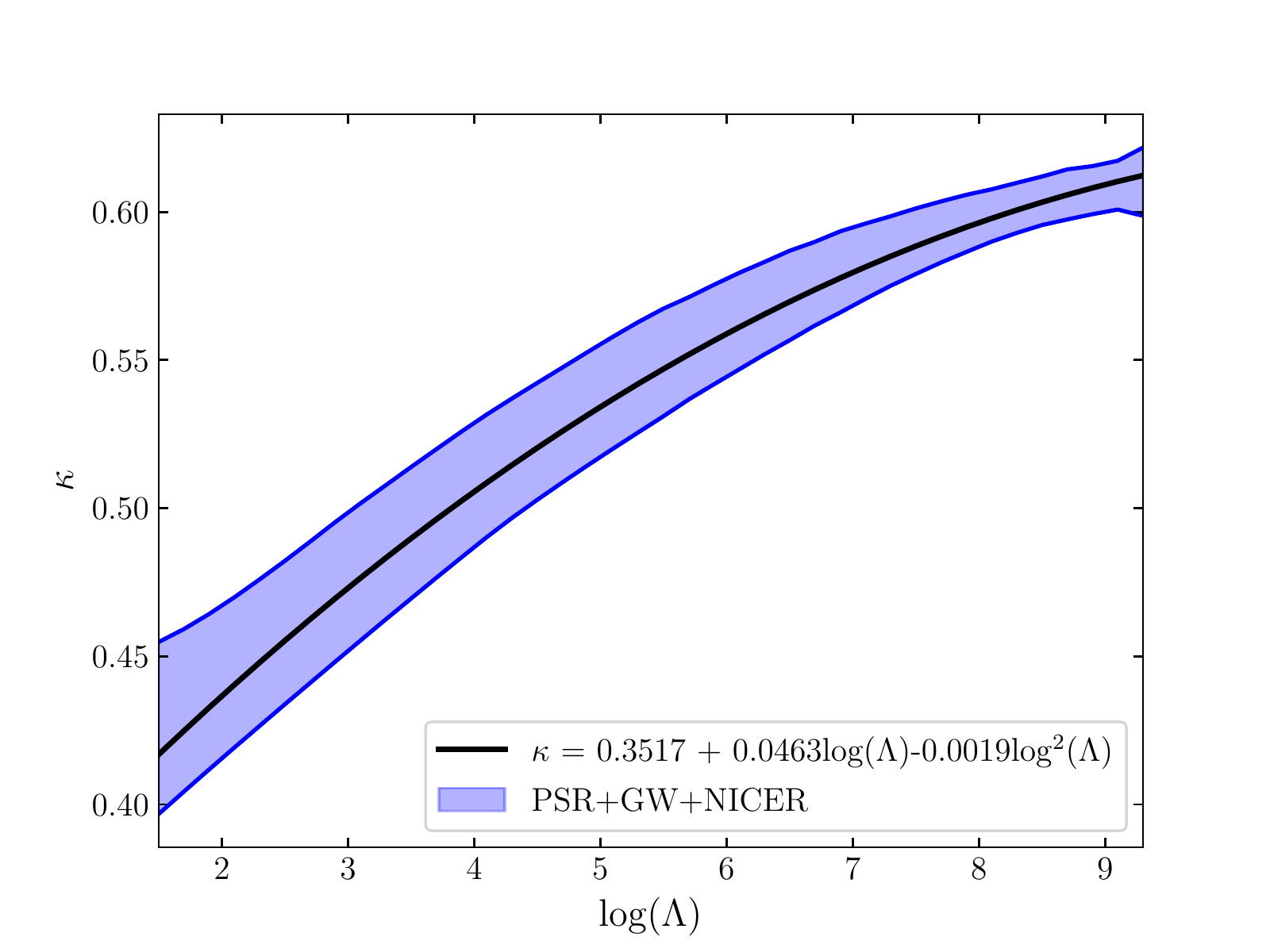}
\end{minipage}
\caption{The upper panel shows the $95\%$ level of the r-mode frequency for a range of compactness and the best quadratic model fit to the posterior. The lower panel shows the same as a function of tidal deformability}
\label{fig:phil_k}
\end{figure}

 Several different parametrizations of the neutron star EOS have been proposed  to constrain the EOS from multi-messenger observations of the neutron stars.  Generic parametrizations in terms of piecewise polytropes~\citep{Annala,Hebeler2013,Read2009,Gamba2019},  spectral  decomposition~\citep{Fasano2019,Lindblom2018} and speed-of-sound~\citep{Tews_2018,Greif2019} have been extensively used  for such studies. Here we consider  the  nonparametric  representation constructed through Gaussian process~\citep{Landry2019,Essick2020} rather than a parametrization for the EOS. This  EOS representation allows more model
freedom and can account for different degrees of freedom. In ~\cite{Legred2021}, using this model they studied the implications of the following multi-messenger observations for the neutron star equation of state (EOS)-
 \begin{itemize}
     \item the radio mass measurements for J0348+0432~\citep{J0348} and J0740+6620~\citep{Cromartie,fonseca2021refined}.
     \item the GW mass and tidal deformability measurementsfrom GW170817~\citep{Abbott2017,Abbott2018,Abbott2019} and GW190425~\citep{GW190425}
     \item  mass and radius constraints from NICER observations of J0030+0451~\citep{NICER0030_Miller,NICER0030_Riley} and J0740+6620~\citep{NICER0740_Miller,NICER0740_Riley}
 \end{itemize}
 Using these combined radio, GW and x-ray data, they put constraints for the microscopic EOS and the macroscopic NS properties, masses, radius and tidal deformabilities. In Fig~\ref{fig:phil_eos}, we plot the $95\%$ confidence interval for the pressure-energy density relation using the posterior obtained by combining all the constraints~\citep{legredisaac2022}.
From this posterior set of 10,000 EOSs, we choose 1000 EOS randomly and calculate the r-mode frequency as a function of compactness and tidal deformability. In Fig.~\ref{fig:phil_k}, we plot the $95\%$ confidence interval for the r-mode frequency $\kappa$ as a function of compactness and tidal deformability. We find a substantial spread unlike the Fig.~\ref{fig:f_tabular} for tabulated EOSs owing to the fact that we now have considered 1000 EOSs which span a wide region in the pressure-density relation~\ref{fig:phil_eos}.  

The best fits for the r-mode frquency $\kappa$ to our quadratic models are given by
\begin{equation}\label{eq:univ_phil}
    \kappa = 0.67 - 0.593C - 0.772C^2,
\end{equation}
\begin{equation}\label{eq:univ_phil_lamb}
    \kappa = 0.3517 + 0.0463 \> log(\Lambda) - 0.0019 \> log^2(\Lambda).
\end{equation}
 
\section{Discussion and Conclusion}
\label{sec:discussion}

In this work, we solved perturbation equations for relativistic barotropic stars consistently with the  boundary conditions using a spectral method and calculated the r-mode frequency as a function of compactness for various equation of states. First we checked the convergence of our code and then compared our results for $n=0$ polytrope with Fig. 1 from ~\cite{Lockitch2003} and for $n=1$ polytropes with Fig. 2 from ~\cite{idrisy2015}. We see that our results match up to $0.1\%$ for $n=0$ polytrope from~\cite{Lockitch2003}. Since the precision of the converged $\kappa$ value in our case is upto 3 decimal places, an error of the order $\leq 1\%$ is expected. But we see a greater deviation $\sim 2-3\%$ for the results for $n=1$ polytrope from~\cite{idrisy2015}. Probable reason for this much larger deviation could be an erroneous representation of the perturbation and boundary condition (Eq. 41 and Eq. 50) in~\cite{idrisy2015} in comparison with Eq. 20 and Eq. A4 in~\cite{Lockitch2003} respectively. From Table~\ref{tab:eos} we find that, while considering the realistic EOS tables, there is also around $3{\%}$ deviation in the r-mode frequency from~\cite{idrisy2015}. \\

To get the frequency band to look for r-modes from astrophysical sources, we should consider the spread of compactness in the neutron star population. The compactness is estimated by the ratio of its stellar mass to the radius~\citep{idrisy2015}
\begin{equation}\label{eq:compact}
    \frac{M}{R} \approx 0.207\left(\frac{M}{1.4M_{\odot}}\right)\left(\frac{10\text{ km}}{R}\right).
\end{equation}
The minimum value of neutron star mass is taken to be 1$M_{\odot}$ from Fig 1 of~\cite{Lattimer2010}. The limit is conservative in a sense that stellar cores with lesser mass probably would not go through supernova explosion to produce neutron stars. For our set of tabulated EOSs listed in Table~\ref{tab:eos}, we find the maximum radius for 1$M_{\odot}$ stars to be around $14.5$km. This gives the lower limit of compactness to be $\approx 0.103$. Assuming the casuality of EOSs, there is a upper limit to the compactness $\frac{M}{R} \leq 0.35$~\citep{LATTIMER2007} but none of the EOSs from Table~\ref{tab:eos} reach this high value for any stable configuration. ~\cite{idrisy2015} uses the compactness range $0.11 - 0.31$, but we give our limits of the r-mode frequency for both compactness ranges in Table~\ref{tab:limits}. \\

Now, to calculate the narrow frequency band to search over in LIGO data, the r-mode frequency($f$) is given in terms of the rotational frequency($\nu$)~\citep{Caride2019,LIGO_rmode} as
\begin{equation}
    \frac{f}{\nu} = A - B\left(\frac{\nu}{\nu_K}\right)^2,
\end{equation}
where $\nu_K$ is the Keplerian frequency = 506 Hz considered in ~\cite{LIGO_rmode}. The uncertainties in the range of $A$ and $B$ give the parameter space for the GW signal model~\citep{Caride2019, LIGO_rmode}. General relativistic corrections for slowly rotating stars give the range of A~\citep{idrisy2015} and rapid rotation correction gives the range of B~\citep{yoshida2005}. The ranges are given by ~\citep{Caride2019} 
\begin{align}\label{eq:range}
     1.39 &\leq A \leq 1.57, \nonumber \\
    0 &\leq B \leq 0.195~.
\end{align}
Since, the r-mode frequency in the inertial frame is given by $f = |(\kappa - m)\Omega|$, we can use the limits on $\kappa$ to update the limits on $A$ value which is also listed in Table~\ref{tab:limits}. Considering all these multi-messenger constraints and a conservative limit on the possible ranges of compactness, we put a limit on the value of A to be $1.39 \leq A \leq 1.64$ which should be used for narrow-band gravitational wave searches for known pulsars.

\begin{table}[ht]
\caption{Proposed ranges of $\kappa$ and $A$~\citep{Caride2019} for ranges of compactness comparing universal relations from ~\cite{idrisy2015}  with our results for both Tabulated EOS and Non-parametric EOSs}
\centering
\begin{tabular}{||c |c |c ||} 
 \hline
Universal relation  & $\kappa$ &  $A$  \\ [0.5ex] 
 (compactness range) & & \\
 \hline\hline
 ~\citep{idrisy2015}(0.11-0.31) & 0.614-0.433 & 1.39-1.57  \\
 \hline
 Tabulated EOS\eqref{univ_real}(0.11-0.31) & 0.601-0.412 & 1.40-1.59 \\
 \hline
 Tabulated EOS\eqref{univ_real}(0.10-0.35) & 0.608-0.364 & 1.39-1.64 \\
 \hline
 Non-parametric EOS\eqref{eq:univ_phil}(0.11-0.31) & 0.596-0.415 & 1.39-1.59 \\
 \hline
 Non-parametric EOS\eqref{eq:univ_phil}(0.10-0.35) & 0.604-0.371 & 1.40-1.63\\
 \hline
\end{tabular}
\label{tab:limits}
\end{table} 

 PSR J0537-6910 is particularly interesting for r-mode searches because it is the fastest-spinning known  young  pulsar  with  rotation  frequency $\nu$ =62 Hz which places gravitational-wave frequency in the LIGO sensitivity band and its inter-glitch braking index $n \approx 7$ which is expected for GW emission via r-mode~\citep{Anderrson2018}. In the latest LIGO r-mode search from PSR J0537, values of $A$ and $B$ given in~\eqref{eq:range} gives the r-mode frequency band 86-98 Hz. If we use our universal relation~\eqref{univ_real} for a compactness range of $0.11-0.31$ (same as in ~\cite{idrisy2015}), we get the frequency band 87-99 Hz which is 1 Hz higher in both lower and upper bounds. This might not appear very large but since we are looking at narrow frequency band searches for these analyses, this new universal relation introduces a $8.3\%$ change (1 Hz in a frequency band of 12 Hz) in the frequency band. For a compactness range of $0.10-0.35$, this range becomes 86-101 Hz. \\

We also provided universal relations for the r-mode frequency $\kappa$ and the dimsionless tidal deformability($\Lambda$) in Eq.~\eqref{eq:univ_lamb} and Eq.~\eqref{eq:univ_phil_lamb}. Recent work by ~\cite{pawan2022} used such a relation to reconstruct the EOS with observations of the inspiral signal by Einstein Telescope (ET) with or without r-modes. They used the universal relation for $\kappa$ vs compactness from~\cite{idrisy2015} to derive the r-mode frequency as a function of $log(\Lambda)$. If we compare with Eq. (7) from~\cite{pawan2022} against our universal relation, we observe a $5-6\%$ difference in the r-mode frequency which can be significant while constraining the nuclear EOS from inspiral signal with r modes. \\

To conclude, we have derived the r-mode frequency as a function of compactness for neutron stars and showed that, one can obtain a universal relation between r-mode frequency and the compactness which is independent of the EOSs. With a physically motivated range of compactness, we derived the frequency band to search for r-modes in the LIGO data. For the particular interesting candidate PSR J0537-6910, we showed that our narrow-band frequency range can vary from the previous searches by $8-25\%$ depending on the compactness range chosen. If a continuous wave from r-mode is detected, these universal relations can also be used to constrain the nuclear EOS but distinguishing between different EOSs will be difficult as the deviation in r-mode frequency is $< 1\%$ for different EOSs. Using ET, if r-mode excitation is observed in the inspiral signal, we also can constrain the tidal deformability and the EOS using our universal relations. Although in this study, we ignored rapid rotation and other physical mechanisms inside the neutron star that can affect the r-mode frequency, they might be important for some particular neutron stars and in future, we would like to explore how these mechanisms affect r-mode frequency for different EOSs.    

\section{Acknowledgements}
The authors would like to thank David Ian Jones for his useful comments and suggestions on the material of this paper. The authors acknowledge usage of the IUCAA HPC computing facility for the numerical calculations. 

\bibliography{sample63}{}
\bibliographystyle{aasjournal}



\end{document}